\documentclass[10pt]{article}
\usepackage{graphicx}

\def\Title#1{\begin{center} {\Large #1 } \end{center}}
\def\Author#1{\begin{center}{ \sc #1} \end{center}}
\def\Address#1{\begin{center}{ \it #1} \end{center}}

\newcommand\pubblock{\rightline{\begin{tabular}{l} Proceedings of the Second Annual LHCP\\ \pubnumber\\
         \pubdate  \end{tabular}}}

\newenvironment{Abstract}{\begin{quotation} \begin{center} 
             \large ABSTRACT \end{center}\bigskip 
      \begin{center}\begin{large}}{\end{large}\end{center} \end{quotation}}

\newenvironment{Presented}{\begin{quotation} \begin{center} 
             PRESENTED AT\end{center}\bigskip 
      \begin{center}\begin{large}}{\end{large}\end{center} \end{quotation}}

\def\Acknowledgements{\bigskip  \bigskip \begin{center} \begin{large}
             \bf ACKNOWLEDGEMENTS \end{large}\end{center}}

\def\beq{\begin{equation}}
\def\eeq#1{\label{#1}\end{equation}}
\def\eeqn{\end{equation}}

\def\beqa{\begin{eqnarray}}
\def\eeqa#1{\label{#1}\end{eqnarray}}
\def\eeqan{\end{eqnarray}}

\let\bar=\overbar

\def\ie{{\it i.e.}}
\def\eg{{\it e.g.}}

\def\Dslash{\not{\hbox{\kern-4pt $D$}}}
\def\dslash{\not{\hbox{\kern-2pt $\del$}}}

\def\msb{{\bar{\ssstyle M \kern -1pt S}}}

\usepackage{ifthen} 
\newboolean{pdflatex}
\setboolean{pdflatex}{true} 
 
\newboolean{articletitles}
\setboolean{articletitles}{true} 
 
\newboolean{uprightparticles}
\setboolean{uprightparticles}{false} 
 
\newboolean{inbibliography}
\setboolean{inbibliography}{false} 

\usepackage{color}
\usepackage{cite}
\usepackage{mciteplus}
\usepackage{hyperref}

\newcommand\pubnumber{ }

\newcommand\pubdate{\today}

\def\affiliation{
On behalf of the LHCb collaboration, \\
CERN-PH \\
CH-1211 Geneva 23, Switzerland}

\begin{document}
\newcommand{\jpsi} {\ensuremath{J\mskip -2.5mu/\mskip -1mu\psi\mskip 1mu}}
\newcommand{\psitwos}  {\ensuremath{\psi(2S)}}
\newcommand{\lum}{\ensuremath{\mathcal{L}}}
\newcommand{\tev}{\ensuremath{\,\mathrm{TeV}}}
\newcommand{\gevc}{\ensuremath{\,\mathrm{GeV}\mskip -2mu/\mskip -1mu c}}
\newcommand{\mevc}{\ensuremath{\,\mathrm{MeV}\mskip -2mu/\mskip -1mu c}}
\newcommand{\gevcc}{\ensuremath{\,\mathrm{GeV}\mskip -2mu/\mskip -1mu c^2}}
\newcommand{\mevcc}{\ensuremath{\,\mathrm{MeV}\mskip -2mu/\mskip -1mu c^2}}
\newcommand{\pbinv}{\ensuremath{\,\mathrm{pb}^{-1}}}
\newcommand{\nbinv}{\ensuremath{\,\mathrm{nb}^{-1}}}
\newcommand{\fbinv}{\ensuremath{\,\mathrm{fb}^{-1}}}
\newcommand{\pt}{\ensuremath{p_{\rm T}}}
\newcommand{\microb}{\ensuremath{\,\upmu\mathrm{b}}}

\large
\begin{titlepage}
\pubblock

\vfill
\Title{ Exotic charmonium-like spectroscopy  at LHCb: \\
a study of the  X(3872) and of the Z(4430)$^-$ states }
\vfill

\Author{Monica Pepe Altarelli}
\Address{\affiliation}
\vfill
\begin{Abstract}

I will report on the recent LHCb results on the evidence for the decay $X(3872)\to\psitwos\gamma$, and on the improved measurement of the mass and width of the  $Z(4430)^-$, the determination of its quantum numbers and the observation of its resonant character.

\end{Abstract}
\vfill

\begin{Presented}
The Second Annual Conference\\
 on Large Hadron Collider Physics \\
Columbia University, New York, U.S.A \\ 
June 2-7, 2014
\end{Presented}
\vfill
\end{titlepage}
\def\thefootnote{\fnsymbol{footnote}}
\setcounter{footnote}{0}
%

\normalsize 


\section{Introduction}
In the na{\"i}ve quark model, only two types of quark combinations are required to account for the existing hadrons: $q\bar{q}$ combinations form mesons, while baryons are made up of three quarks. However, the last few years have seen the discovery of a large number of states, generically denoted as $X$$, Y$ and $Z$, that do not fit the conventional picture, including many unexpected neutral and several charged ones. The existence of $\eg$ tetraquarks ($q\bar{q}q\bar{q}$), hadronic molecules, and other bound states involving gluons ($gg$ and $q\bar{q}g)$ has been invoked to explain some of these exotic states, but a compelling unified description has not yet emerged. 
The LHCb experiment has already collected an impressive catalogue of results on new and excited heavy flavour states in its first two years of operation, as well as first time observations of many new decay modes.  With its large statistics, efficient trigger, excellent momentum resolution and particle identification, it is an ideal laboratory to perform hadron spectroscopy studies with high precision.
\section{$ X(3872)$}
The $X(3872)$, discovered by Belle in 2003~\cite{Choi:2003ue} as an unexpected structure in the $\jpsi\pi^+\pi^-$ invariant mass while studying $B^+\rightarrow K^+\pi^+\pi^-\jpsi$ decays,  was the first charmonium-like state found not to fit a conventional quarkonium description.
The $X(3872)$ is a particularly intriguing state because on the one hand the fact that it decays into $\jpsi\pi^+\pi^-$ leads to a natural interpretation as a charmonium excitation; on the other hand the closeness of its mass  to the $D^{*0}\bar{D^{ 0}}$ threshold ($\simeq 3872\,\rm{MeV}$) and its prominent decays to $D^{*0}\bar{D^{ 0}}$~\cite{PDG2012} suggest that it may be an example of a hadron molecule with an extremely small binding energy (measured to be $0.09\pm0.28\,\rm{MeV}$ in Ref.~\cite{Aaij:2013uaa}). Moreover, the decay modes $X(3872)\rightarrow\jpsi \rho$ and $X(3872)\rightarrow\jpsi \omega$ were observed to have comparable branching fractions, revealing a severe violation of isospin symmetry. This observation is compatible with a molecular interpretation of the $X(3872)$ because the $D^{*0}\overline{D^{ 0}}$ state is a superposition of isospin 0 and 1, while  the $D^{*+}{D^{ -}}$ threshold is about 8~MeV higher. 

After measuring the $X(3872)$ mass and production cross-section in $pp$ collisions at $\sqrt{s}=7\tev$~\cite{Aaij:2011sn}, LHCb has more recently performed a full five-dimensional angular analysis of the $\jpsi\pi^+\pi^-$ decays of the $X(3872)$, produced in $B^{+} \rightarrow X(3872)K^{+}$~\cite{LHCb-PAPER-2013-001}. In this analysis the quantum numbers of the $X(3872)$ are determined to be  $J^{PC}=1^{++}$, while the only alternative assignment allowed by previous measurements $J^{PC}=2^{-+}$~\cite{Abulencia:2006ma} is ruled out with a significance of more than eight standard deviations.  The $1^{++}$ quantum numbers are those of the conventional  charmonium state $\chi_{c1}(2P)$, which is however disfavoured by the value of the $X(3872)$ mass, leaving open the possibility of more unconventional interpretations. 

To study this further, LHCb has recently measured~\cite{LHCb-PAPER-2014-008} the ratio of branching fractions
\beq
 R_{\psi\gamma}=\frac{B(X(3872)\rightarrow\psi(2S)\gamma)}{B(X(3872)\rightarrow \jpsi\gamma)}\, ,
 \eeqn
 as a constraint on the charmonium content of the $X(3872)$. The branching fraction $B(X(3872)\rightarrow\psi(2S)\gamma)$ is in fact expected to be very small for a pure molecule $(O(10^{-3}))$~\cite{Swanson:2004cq,Dong:2009uf,Ferretti:2014xqa}, but it could be enhanced for an admixture of a $D^{*0}\overline{D^{ 0}}$ molecule and charmonium. The BaBar collaboration has measured a relative large branching fraction for the $X(3872)$ into $\psi(2S)\gamma$, with $R_{\psi\gamma}=3.4\pm1.4$~\cite{Aubert:2008ae}, a result generally inconsistent with a pure molecular interpretation; in contrast, no significant signal was found by Belle~\cite{Bhardwaj:2011dj}.
 
\begin{figure}[htb]
\centering

\includegraphics[height=3.0in]{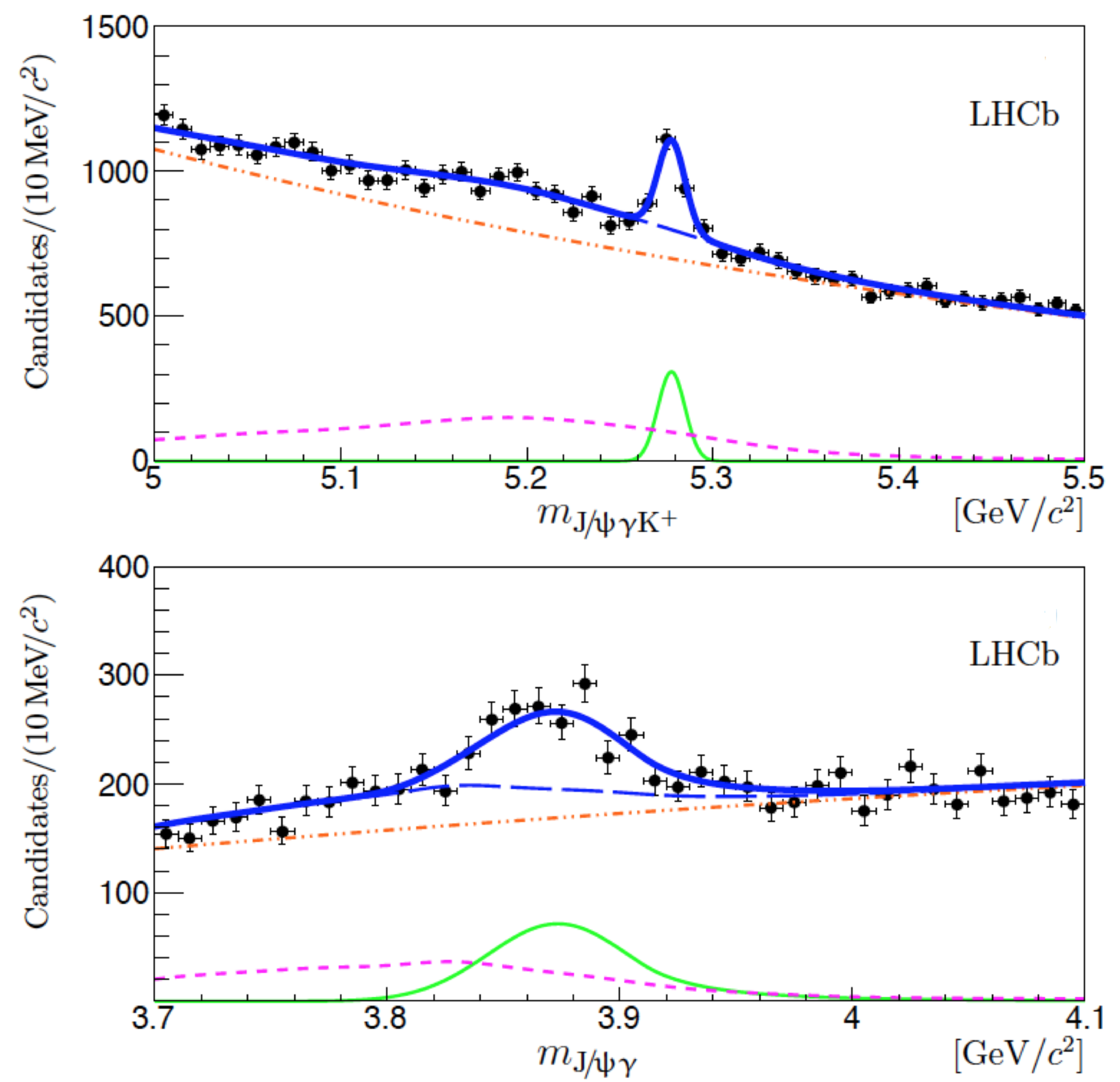}
\includegraphics[height=3.0in]{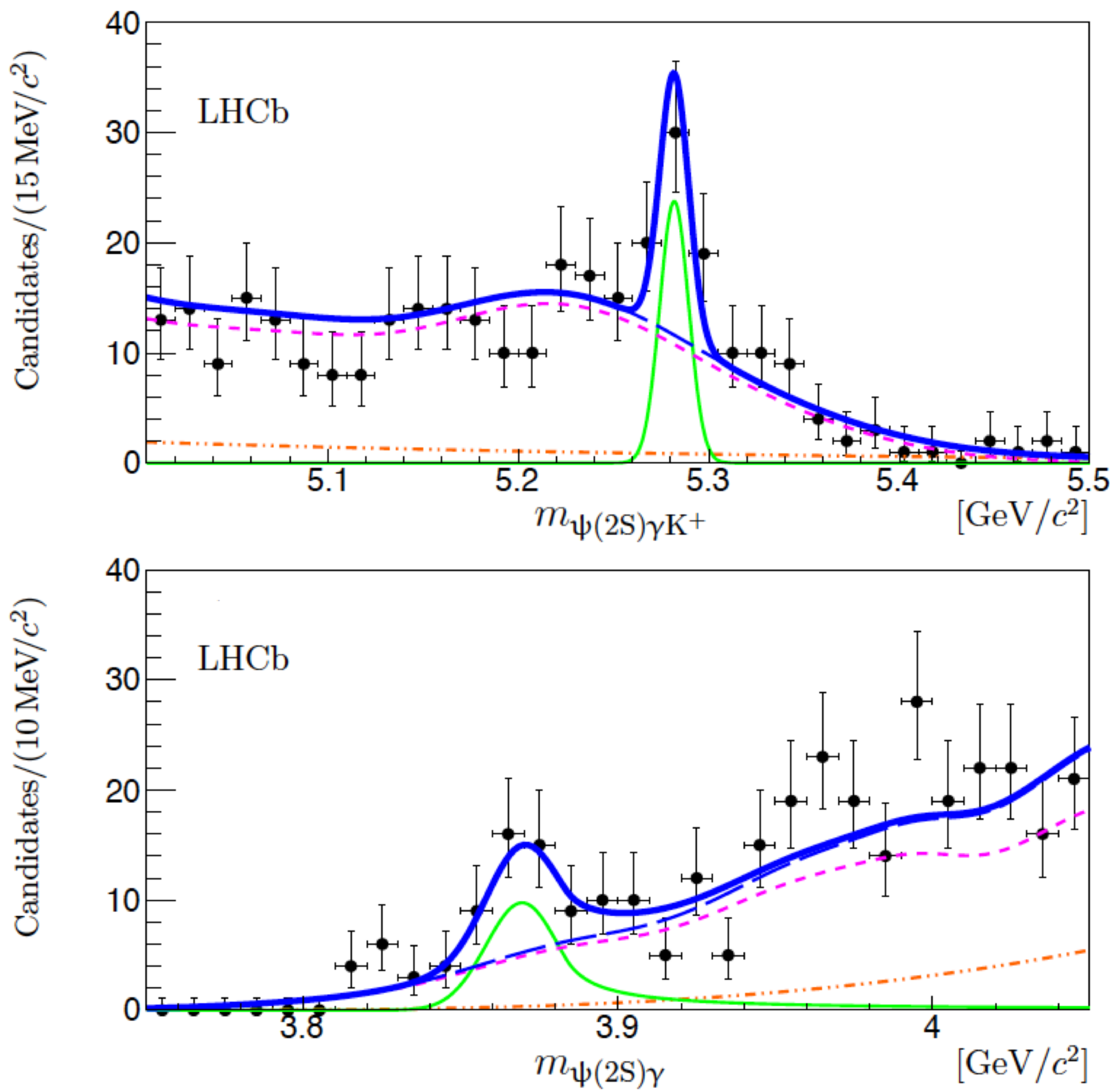}
\caption{ Invariant mass distribution for $B^{+}\rightarrow X(3872)K^+$ candidates for the $J/\psi$ (top left) and   $\psi(2S)$ (top right) channels, restricted to those  with $J/\psi\gamma$ or $\psi(2S)\gamma$invariant mass within $\pm 3\sigma$ from the $X(3872)$ peak position. Invariant mass distribution for  $\jpsi\gamma$ (bottom left) and $\psi(2S)\gamma$ (bottom right) candidates, restricted to those  with $\jpsi\gamma K^{+}$ or $\psi(2S)\gamma K^{+}$ invariant mass within $\pm 3\sigma$ from the $B^{+}$ peak position. The total fit (thick solid blue) together with the signal (thin solid green) and background components (dash-dotted orange for the combinatorial, dashed magenta for the peaking component and long-dashed blue for their sum) are shown. }
\label{fig:figure1}
\end{figure}

In LHCb a search for the $X(3872)$ into $\psi(2S)\gamma$ is performed by using $B^{\pm}\rightarrow X(3872)K^{\pm}$ decays and reconstructing the $\psi(2S)$ meson in the $\mu^+\mu^-$ channel. The analysis is based on a data sample of $pp$ collisions corresponding to an integrated luminosity of 1~fb$^{-1}$ at $\sqrt{s}=7\tev$ and 2~fb$^{-1}$ at $\sqrt{s}=8\tev$, $\ie$, the full LHCb Run 1 data sample. Figure~\ref{fig:figure1} shows the invariant mass distribution of the $B^{\pm}\rightarrow X(3872)K^{\pm}$ candidates for the $J/\psi$ (top left) and   $\psi(2S)$ (top right) channels, while the bottom-left and bottom-right plots display  the $\jpsi\gamma $ and $\psi(2S)\gamma$  invariant mass distributions, respectively. The observed signal in the $\psi(2S)$ channel is $36.4\pm9.0$ events; the significance for such a signal  is determined to be 4.4 standard deviations by simulating a large number of background-only experiments. From the measured yields, the ratio of branching fractions is computed to be $ R_{\psi\gamma}=2.46\pm0.64\pm0.29$, where the first uncertainty is statistical and the second is systematic. This result, which improves with respect to previous measurements~\cite{Aubert:2008ae, Bhardwaj:2011dj}, favours the interpretation of the $X(3872)$ as an admixture of a $D^{*0}\overline{D^{ 0}}$ molecule and charmonium, as opposed to a pure molecular interpretation.

\section{$Z(4430)^-$}

The $Z(4430)^-$ state was observed by the Belle collaboration~\cite{Choi:2007wga, Mizuk:2009da} as a charged resonance structure in the $\psi(2S)\pi^-$ invariant mass distribution of the decay $B^{-,0}\rightarrow\psi(2S)K\pi^-$ (with $K\,=\,K^0_s$ or $K^+$). Such a resonance, decaying to a heavy quarkonium state  and a charged light-quark meson, is particularly interesting  as it is most likely exotic, with minimal quark content $c\bar{c}u\bar{d}$. The BaBar collaboration searched for evidence to support the existence of the $Z(4430)^-$ in the $J/\psi\,\pi^-$ or $\psi(2S)\pi^-$ mass distributions of the decays  $B^{-,0}\rightarrow\psi(2S)K\pi^-$ and $B^{-,0}\rightarrow J/\psi K\pi^-$~\cite{Aubert:2008aa}. They performed a model-independent analysis and found that the $J/\psi\,\pi^-$ and $\psi(2S)\pi^-$ mass distributions could each be well described by reflections of known $K^*$ systems with spin $J\le 3$ without invoking exotic states; however, they could not rule out the Belle results. Belle's latest result~\cite{Chilikin:2013tch}, based on a full amplitude analysis, gives a significance of 5.2 standard deviations for the $Z(4430)^-$ state;  they measure  the $Z(4430)^-$ mass to be $M_{Z^-}=4485\pm22^{+28}_{-11}\,\rm{MeV}$ and width $\Gamma_{Z^-}=200^{+41+26}_{-46-35}\,\rm{MeV}$, and favour the $J^P=1^+$ spin-parity assignment by more than 3.4 standard deviations.

\begin{figure}[htb]
\centering
\includegraphics[height=2.0in]{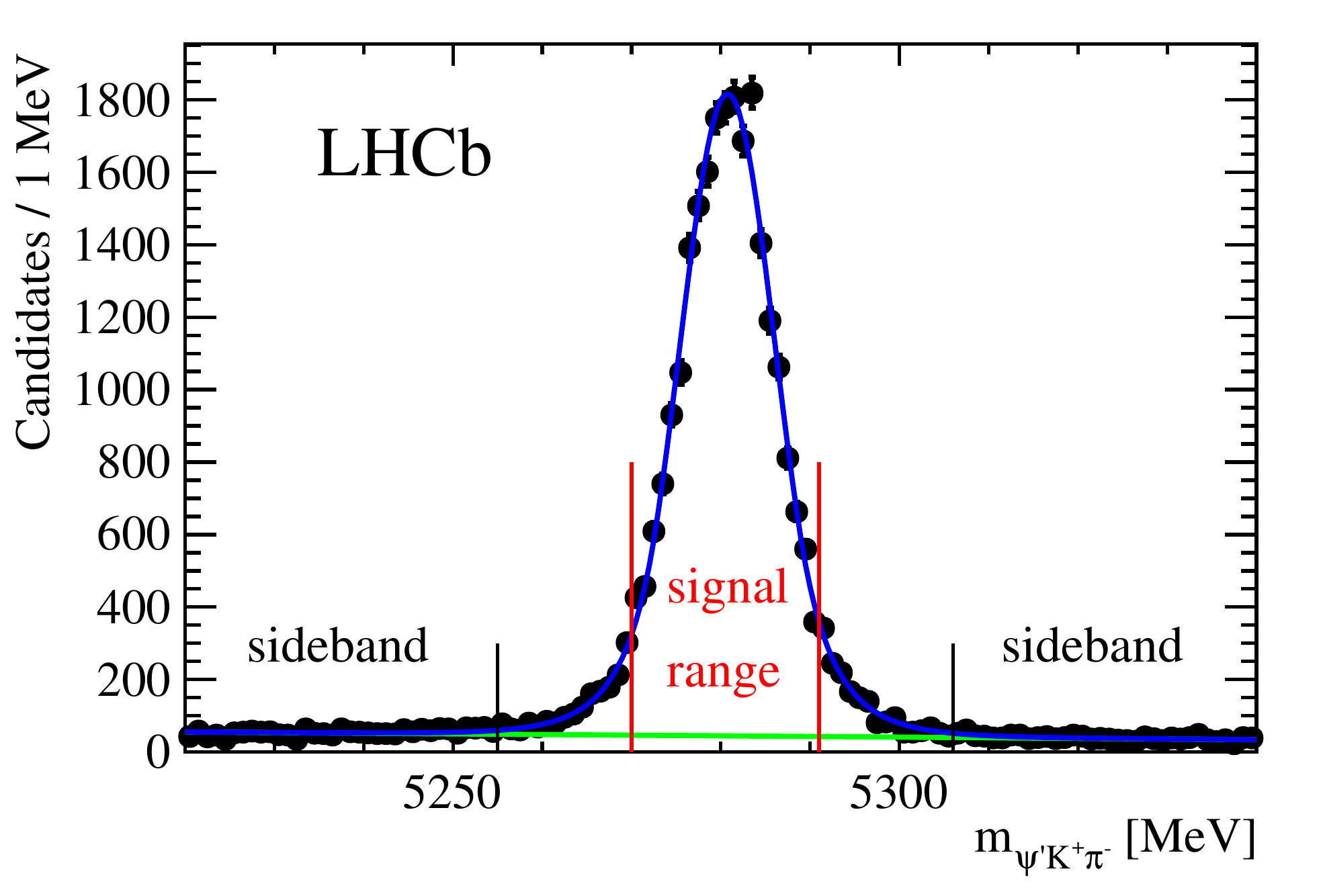}
\includegraphics[height=2.0in]{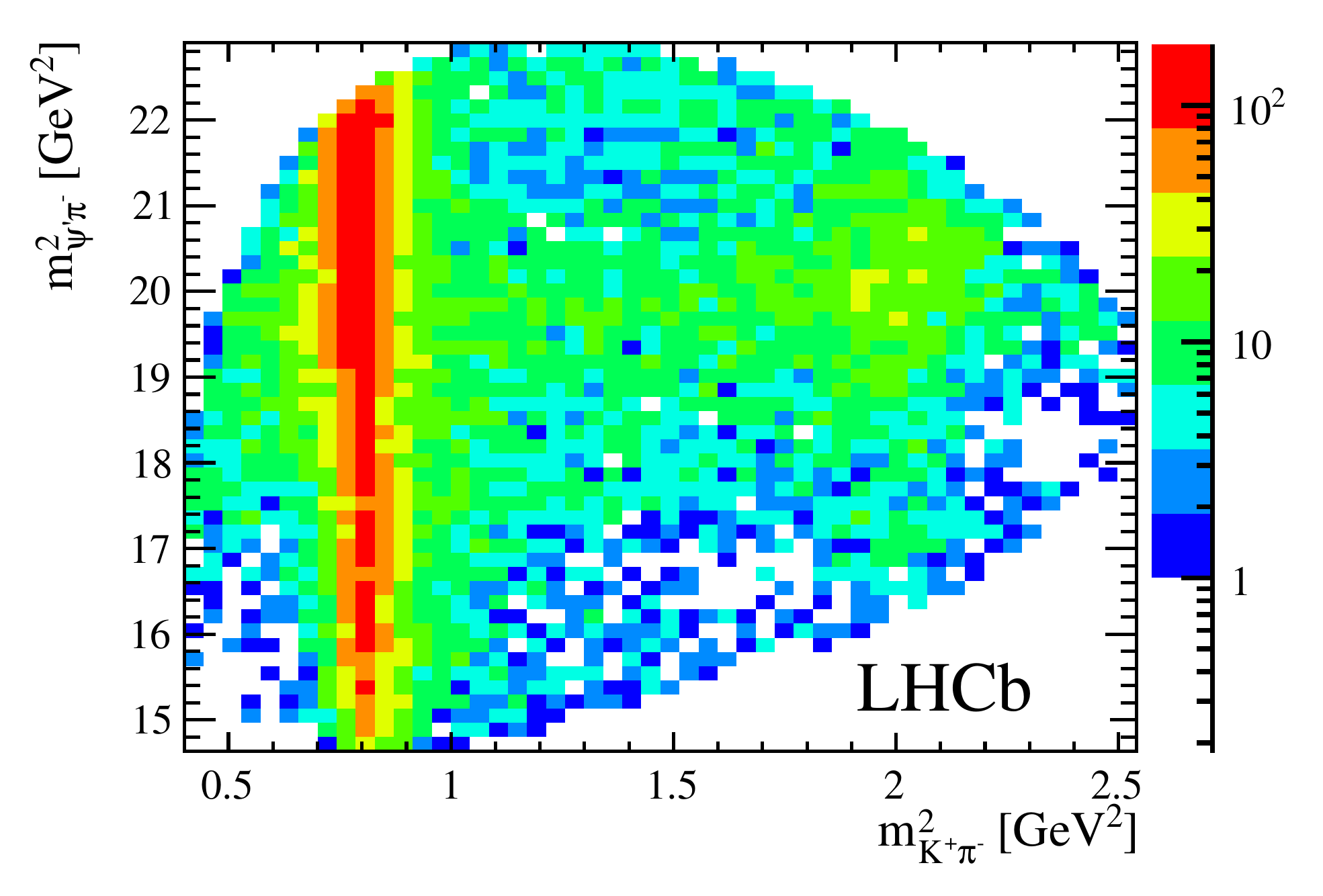}
\caption{ Invariant mass distribution (left) and Dalitz plot (right) for $B^0\rightarrow \psi(2S) K^+\pi^-$ decay candidates. In the invariant mass distribution the fit (blue line) with a double-sided Crystal Ball signal shape and linear background (green line) are superimposed to the data points. }
\label{fig:figure2}
\end{figure}

The LHCb collaboration investigated resonant structures in $B^0\rightarrow \psi(2S) K^+\pi^-$ decays, with the $\psi(2S)$ decaying into two muons, using $pp$ collision data corresponding to an integrated luminosity of $3 \fbinv$~\cite{Aaij:2014jqa}. Some 25,000 $B^0$ candidates were reconstructed with approximately $4\%$ background in the signal region, as shown in Fig.~\ref{fig:figure2}  (left). This represents a ten-fold increase in signal yield over the previous measurements, with  a background reduced by approximately a factor of two. Figure~\ref{fig:figure2}  (right) shows the Dalitz plot for the background-subtracted data;  clear contributions from the  $K^*(892)$ and from the $K^*_2(1430)$ are visible as vertical bands, while the $Z(4430)^-$ appears as a horizontal band at  $m^2_{\psi(2S)\pi^-}\sim20\rm\,{GeV}^2$.

\begin{figure}[htb]
\centering
\includegraphics[height=2.0in]{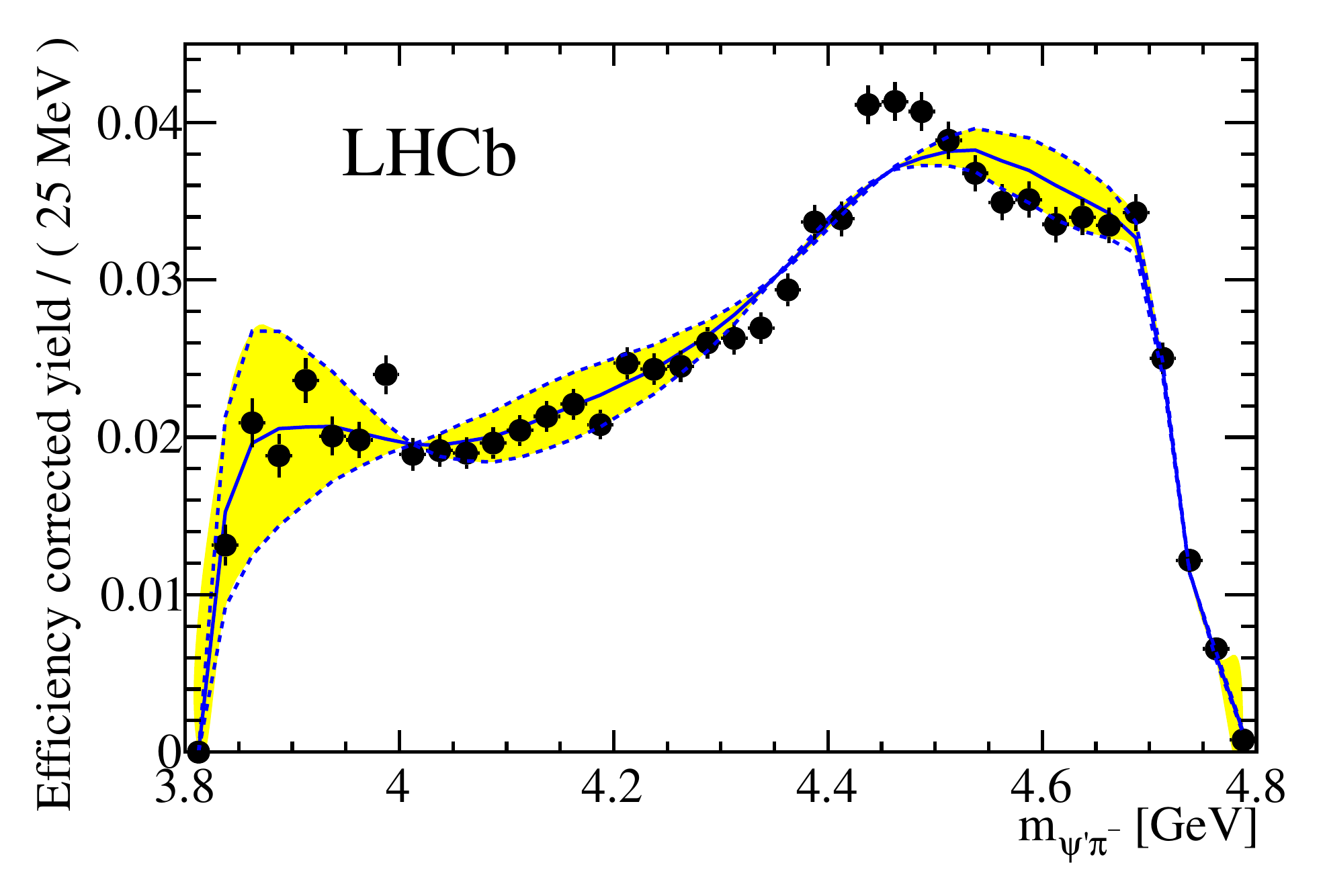}
\caption{ Background-subtracted and efficiency-corrected $m_{\psi(2S)\pi^-}$ distribution superimposed with the reflections of Legendre polynomial moments  and their correlated statistical uncertainty (yellow band bounded by blue dashed lines). }
\label{fig:figure3}
\end{figure}

LHCb has performed an analysis based on the model-independent approach developed by BaBar~\cite{Aubert:2008aa}  to check whether the $m_{\psi(2S)\pi^-}$ spectrum in $B^0\rightarrow \psi(2S) K^+\pi^-$ decays can be understood in terms of any combination of known $K^*$ resonances. No constraint is imposed on these resonances besides restricting their maximal spin to two, as  the $K^+\pi^-$ invariant mass spectrum is dominated by $S$, $P$ and $D$ partial waves. The description of the $K^+\pi^-$ angular structure is performed in terms of Legendre polynomial moments. The analysis shows that the relatively narrow peaking structure in the $m_{\psi(2S)\pi^-}$ distribution at 4440 MeV cannot be described in terms of moments of $K^*$ resonances, as illustrated in Fig.~\ref{fig:figure3}. The inclusion of spin-three contributions does not change this conclusion. 

\begin{figure}[htb]
\centering
\includegraphics[height=2.0in]{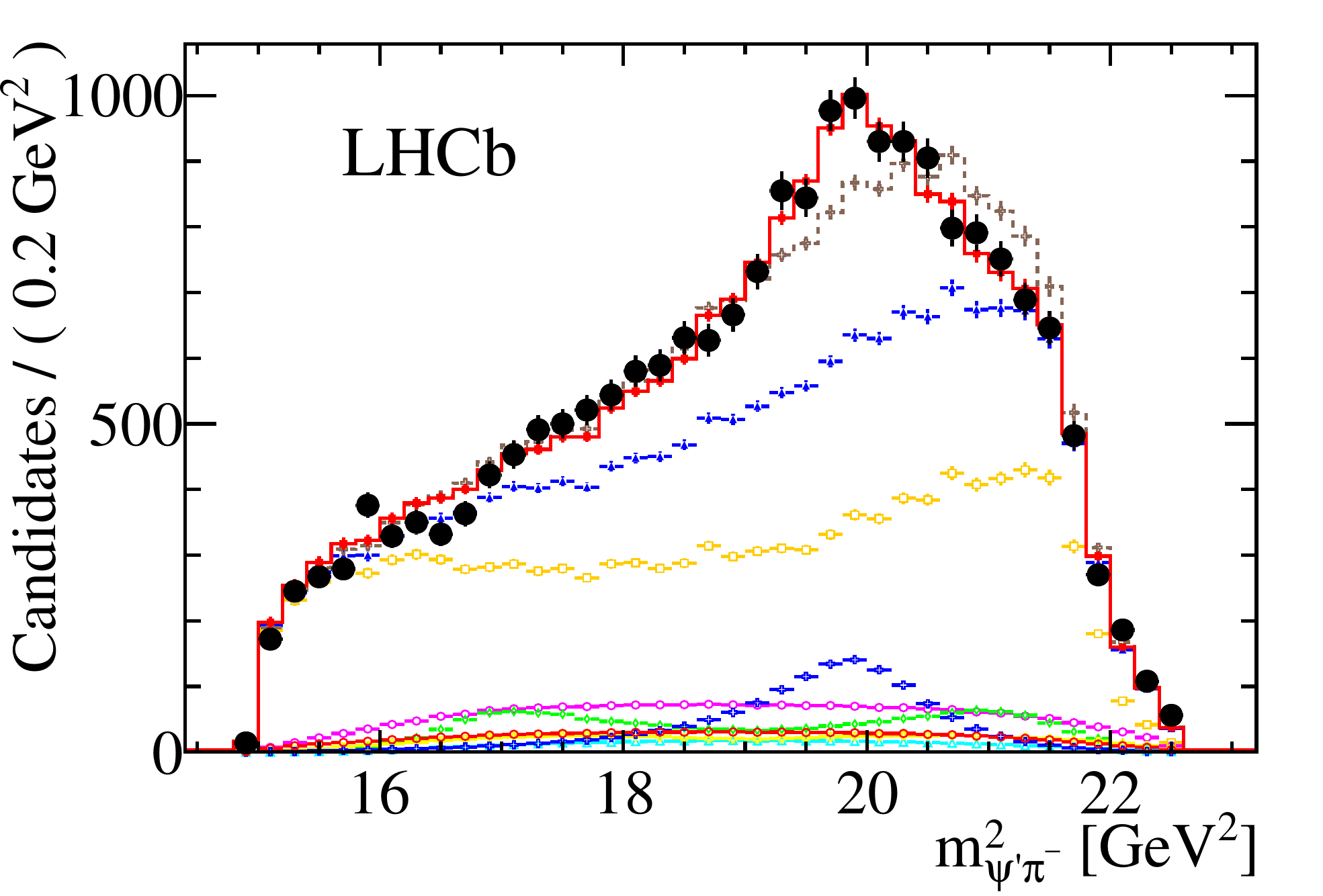}
\includegraphics[height=2.0in]{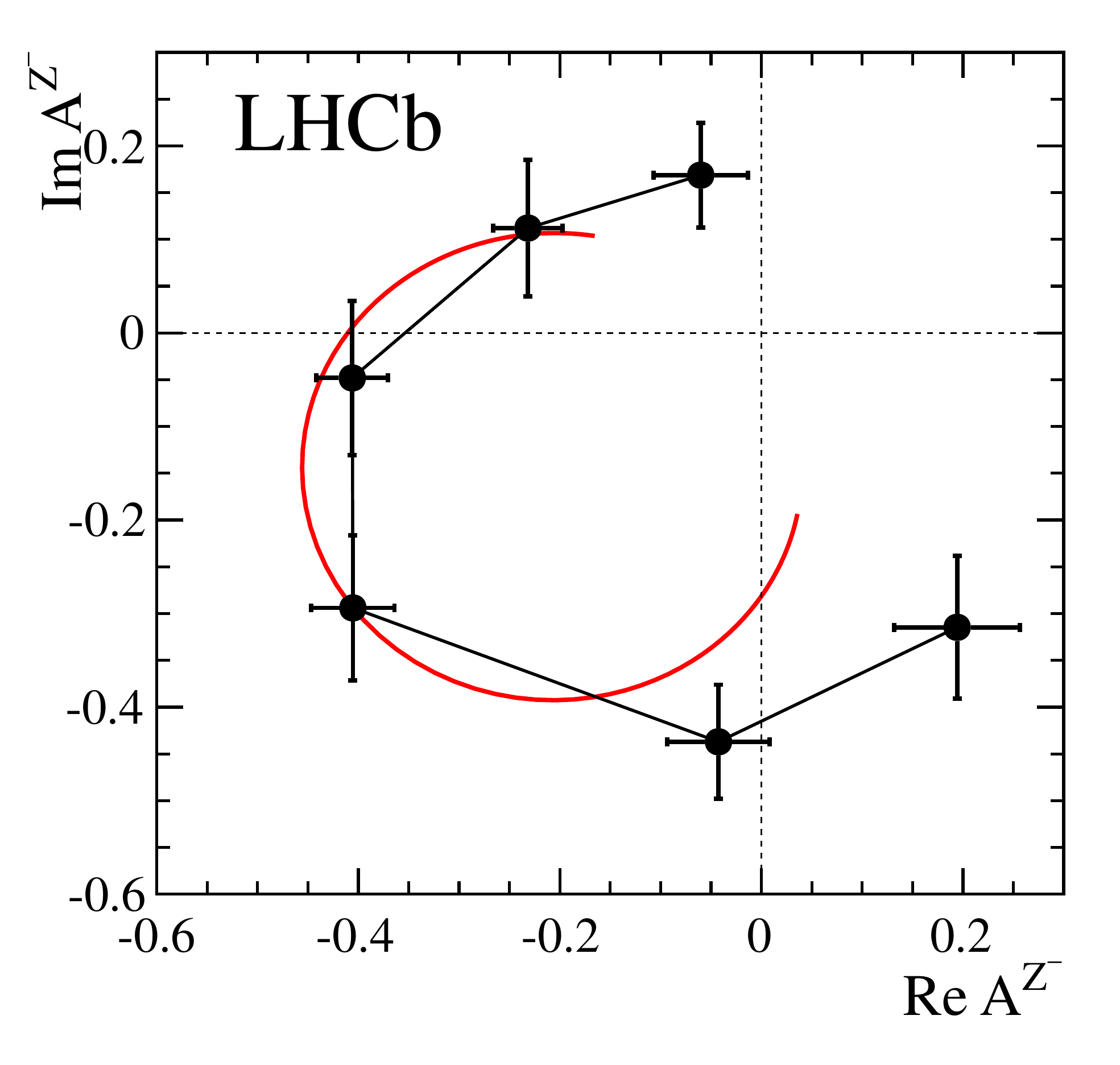}
\caption{ Projection of the data and of the 4D amplitude fit onto the  $m^2_{\psi(2S)\pi^-}$ axis (left). Argand diagram for the fitted binned $Z(4430)^-$ amplitude (right); the Breit-Wigner amplitude evolution with the mass and width from the nominal fit is shown as a red curve. }
\label{fig:figure4}
\end{figure}

A full amplitude fit was performed to be able to extract quantitative information about  the $Z(4430)^-$, such as its mass, width and spin. The amplitude was calculated in a four-dimensional space in the  two invariant masses squared from the Dalitz plot in Fig.~\ref{fig:figure2}  (right)  and $\psi(2S)$ decay angles, following the formalism and notations of Ref.~\cite{Chilikin:2013tch}. The amplitude is represented by a sum of Breit-Wigner contributions including all known $K^+\pi^-$ resonances within or slightly above the kinematic limit. The angle-dependent part of the amplitude is obtained using the helicity formalism. As shown in Fig.~\ref{fig:figure4} (left), the data are not well described when considering only known $K^{*}\rightarrow K^+\pi^-$ resonances, while a significantly better description is obtained when including in the fit a $Z(4430)^-\rightarrow \psi(2S)\pi^-$ with $J^P=1^+$.  Other $J^P$ hypotheses ($0^-,\,1^-,\,2^+$ and $2^-$) are ruled out with a significance larger than nine standard deviations, thus confirming previous indications from Belle. The positive parity assignment rules out the interpretation in terms of rescattering of a pair of mesons $\bar{D}^*(2010)D_1(2420)$ in a relative $S$-wave~\cite{Rosner:2007mu} . The measured mass $M_{Z^-}=4475\pm7^{+15}_{-25}\,\rm{MeV}$, width $\Gamma_{Z^-}=172\pm13^{+37}_{-34}\,\rm{MeV}$ and fraction of the integral of the amplitude squared attributable to the $Z(4430)^-$, $f_{Z^-}=(5.9\pm0.9^{+1.5}_{-3.3})\%$, are fully consistent with the recent Belle determination and much improved. The significance of the $Z(4430)^-$ contribution is 13.9 standard deviations.

To address the question whether the $Z(4430)^-$ is a real bound state that follows a resonant behaviour without being ``forced so'' by the amplitude model, a fit is performed in which the Breit-Wigner amplitude is replaced by six independent complex amplitudes equally spaced in $m_{\psi(2S)\pi^-}$ bins in the $Z(4430)^-$ peak region defined as $18.0-21.5\, \rm{ GeV}^2$. The resulting Argand diagram illustrated in Fig.~\ref{fig:figure4} (right) is consistent with a rapid phase transition at the peak of the amplitude, just as expected for a resonance, providing a strong argument in favour of the resonant character of the $Z(4430)^-$ state.

It should be noted that in the region of $m^2_{\psi(2S)\pi^-}$ around 16 to 17 GeV$^2$, the fit does not fully describe the data. If a second $Z^-$ resonance, which peaks at a lower mass than the $Z(4430)^-$ and has a broader width, is allowed in the amplitude, the $p$-value of the ${\chi}^2$ test improves from 12 to 26\%. However, the current data are not sufficient to fully characterize such a resonance.

\section{Conclusions}
LHCb has used the large data sample available from Run 1 to perform high-precision studies of exotic charmonium-like states. After having determined the $X(3872)$ quantum numbers, the decay $X(3872)\rightarrow\psi(2S)\gamma$ is established with a significance of 4.4 standard deviations. This result favours the interpretation of the $X(3872)$ as an admixture of a $D^{*0}\overline{D^{ 0}}$ molecule and charmonium, as opposed to a pure molecular interpretation. LHCb also provided confirmation of the $Z(4430)^-$ state already observed by Belle and established its spin-parity to be $1^+$ with overwhelming significance. The $Z(4430)^-$ mass, width and amplitude fraction are measured to be consistent with the Belle results, with substantially improved uncertainties. The evolution of the phase of the $Z(4430)^-$ amplitude as a function of $m_{\psi(2S)\pi^-}$ in an Argand diagram is that expected for a resonance, thus providing a strong argument in favour of the resonant character of the $Z(4430)^-$ state. These results as well as many others obtained by LHCb during its first two years of operation demonstrate its excellent capabilities and prospects for hadronic spectroscopy.

\newpage
\Acknowledgements
I would like to warmly thank the organisers and friends of LHCP 2014 for their kind invitation and beautiful hospitality including a memorable dinner cruise. I would also like to thank my LHCb colleagues for providing the material discussed here and, in particular, Marco Pappagallo and Vanya Belyaev  for their careful reading of this article.


\setboolean{inbibliography}{true}
\bibliographystyle{LHCb}
\bibliography{Mybibliography}
%
%
%

\end{document}